\documentclass[aps,amsmath,amssymb,prb,twocolumn,showpacs,floatfix]{revtex4}
\usepackage{bm}
\usepackage{epsfig}

\newcommand{\cH}{{\mathcal H_{\textrm{sw}}}}
\newcommand{\cS}{{\mathcal S}}
\renewcommand{\S}{{\bf S}}
\newcommand{\Sx}{{\tilde S^x}}
\newcommand{\Sy}{{\tilde S^y}}
\newcommand{\Sz}{{\tilde S^z}}

\newcommand{\Sp}{{\tilde S^+}}
\newcommand{\Sm}{{\tilde S^-}}
\newcommand{\Spm}{{\tilde S^{\pm}}}

\newcommand{\bA}{{\bf A}}
\newcommand{\bF}{{\bf F}}
\newcommand{\bK}{{\bf K}}
\newcommand{\bG}{{\bf G}}
\newcommand{\br}{{\bf r}}
\newcommand{\bq}{{\bf q}}
\newcommand{\bk}{{\bf k}}
\newcommand{\bQ}{{\bf Q}}
\newcommand{\bM}{{\bf M}}
\newcommand{\nn}{{\nonumber}}
\newcommand{\ba}{{\bf a}}

\newcommand{\Phid}{\Phi^\dagger}

\newcommand{\Pid}{\Pi^\dagger}

\newcommand{\tPid}{\tilde\Pi^\dagger}

\newcommand{\coma}{condmat/}

\newcommand{\AF}{\textrm{AF}}

\newcommand{\gap}{\textrm{gap}}
\newcommand{\Tc}{T_\textrm{c}}

\newcommand{\bd}{b^\dagger}
\newcommand{\bW}{\mathbf{W}}
\newcommand{\bsigma}{\bm\sigma}
\newcommand{\llangle}{\langle\kern-.25em\langle}
\newcommand{\rrangle}{\rangle\kern-.25em\rangle}
\newcommand{\fin}{\textrm{F}}
\newcommand{\w}{\omega}

\newcommand{\1}{{(1)}}
\newcommand{\2}{{(2)}}

\newcommand{\etal}{\emph{et al.}}

\begin{document}

\title{Spin dynamics of stripes}

\author{Frank Kr\"uger}
\author{Stefan Scheidl}

\affiliation{Institut f\"ur Theoretische Physik, Universit\"at zu
  K\"oln, Z\"ulpicher Str. 77, D-50937 K\"oln, Germany}

\begin{abstract}
  The spin dynamics of stripes in high-temperature superconductors and
  related compounds is studied in the framework of a spin-wave theory
  for a simple spin-only model.  The magnon dispersion relation and
  the magnetic structure factor are calculated for diagonal and
  vertical stripes.  Acoustical as well as optical bands are included
  in the analysis.  The incommensurability and the $\pi$ resonance
  appear as complementary features of the band structure at different
  energy scales.  The dependence of spin-wave velocities and resonance
  frequencies on the stripe spacing and coupling is calculated.  At
  low doping, the resonance frequency is found to scale roughly
  inversely proportional to the stripe spacing.  The favorable
  comparison of the results with experimental data suggests that the
  spin-only model provides a suitable and simple basis for calculating
  and understanding the spin dynamics of stripes.
\end{abstract}

\date{\today}

\maketitle

\section{Introduction}
\label{sec.intro}

The evidence for the formation of stripes in high-temperature
superconductors (HTSC) and related materials increases continuously.
After the theoretical prediction \cite{Zaanen+89,Schulz89,Machida89}
of stripes as a combined charge and spin-density wave phenomenon,
years passed until a broad interest was triggered by experiments on
insulating La$_{2-x}$Sr$_x$NiO$_{4+\delta}$ (LSNO) and superconducting
La$_{2-x}$Sr$_x$CuO$_4$ (LSCO).\cite{Tranquada+94,Tranquada+95} More
recent experimental evidence\cite{Dai+98,Mook+98,Howald+02,Hoffman+02}
for stripes in the paradigmatic HTSCs YBaCuO$_{6+\delta}$ (YBCO) and
Bi$_2$Sr$_2$CaCu$_2$O$_{8+\delta}$ (BSCCO) strengthens the expectation
that stripe formation in doped layered perovskites is quite generic.

In spite of the striking evidence for stripes in HTSCs, the causal
connection between stripe formation and superconductivity still is a
mystery.  It is puzzling that both phenomena coexist and that,
nevertheless, stripes tend to suppress
superconductivity.\cite{Tranquada+97,Khaykovich+02} For this interplay
spin order is more relevant than charge order.  In particular, the
strength of spin fluctuations appears to play a central role.  Static
spin order seem to be much less compatible with superconductivity than
dynamic spin order.

At present, one important open question is to what extent the stripe
model can account for spin fluctuations not only at low energies,
where the incommensurate response is observed, but also over a wider
energy range, including the resonance phenomenon at the
antiferromagnetic wave vector (see Refs. \onlinecite{Bourges99,He+02}
and references therein).  The specific form of the dynamic magnetic
response -- including incommensurability and $\pi$ resonance -- gave
rise to doubts that it could be consistent with the stripe
model.\cite{Bourges+00} On the other hand, there are proposals
\cite{Batista+01} that both features may be rooted in a stripe-like
spin-density wave.

In this paper we complement the spin-wave analysis by Batista
\etal.\cite{Batista+01} There, the emphasis was put on the
incommensurate ratio between the spin spacing $a$ and the stripe
spacing $d$ which gives rise to a continuous excitation spectrum.
However, in many cases of interest, this ratio $p:=d/a$ is very close
to a rational or even integer value.  Within the stripe model one
actually expects that integer values are very stable due to a lock-in
of the superstructure into the atomic structure.  This pinning
mechanism is considered as the origin of the so called `1/8 conundrum'
in the cuprates,\cite{Tranquada+95} i.e., the stability of $p=4$ over
a considerable doping range.  Detailed measurements of the
spin-excitation spectrum are available close to integer $p$: $p=3$ in
LSNO,\cite{Bourges+02} $p=4$ in LSCO,\cite{Lake+99} and
$p=4$\cite{Mook+02} and $p=5$\cite{Bourges+00} in YBCO.

In order to test whether these experiments can be consistent with the
spin-wave excitation spectrum of a stripe model in the simplest and
most transparent case, we therefore examine integer $p$.  Particular
attention is paid to the spin-wave band structure in the vicinity of
the antiferromagnetic wave vector.  While the incommensurability as
zero-frequency response is fixed by the geometry of the model, we
calculate the spin-wave velocities at the incommensurability and the
$\pi$ resonance as dynamic features.  We evaluate the dependence of
these quantities on the incommensurability (respectively, the doping
level) and the exchange coupling across the stripes.  By a
quantitative comparison, we determine the value of the exchange
coupling across the stripes as the only \textit{a priory} unknown
model parameter.  In particular, the dependence of the $\pi$ resonance
on doping is found to be consistent with experiments.

Our course starts in Section \ref{sec.model} with the introduction of
the spin-only model that constitutes the basis of our study.  The
linear spin-wave theory is outlined in Sec.~\ref{sec.SWT}.  In
Sec.~\ref{sec.res} we present numerical results for the magnon
dispersion relation, spin-wave velocities, $\pi$ resonance and the
structure factor.  In Sec.~\ref{sec.disc} the results of our theory
are discussed and compared to experimental data.

\section{Model}
\label{sec.model}

\begin{figure}[h]
\epsfig{figure=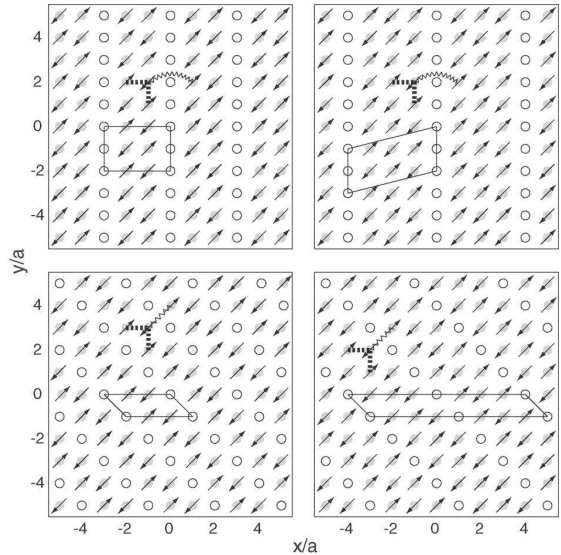,width=0.85\linewidth}
\caption{Illustration of vertical and  diagonal stripe patterns
  with spacings $p=3$ and $p=4$. The hole positions are indicated by
  open circles and the electron positions by grey ones. The arrows
  correspond to the spin orientations in the classical ground state.
  Parallelogramms outline magnetic unit cells spanned by $\bA^\1$ and
  $\bA^\2$.  In our model, we assume antiferromagnetic exchange
  couplings of strength $J$ within the domains (dashed lines) and
  $\lambda J$ across stripes (zig-zag lines).}
  \label{gs} 
\end{figure}

In the cuprates as well as in the nickelates, the metallic spins are
located on square lattices in weakly coupled layers.  Since the the
interlayer coupling generally is much smaller than the intralayer
coupling, we focus on a single layer.  For simplicity, the holes
induced by doping are assumed to form site-centered rivers that act
like antiphase boundaries for the antiferromagnetic
domains.\cite{Tranquada+95} The rivers are assumed to be only one
lattice spacing wide (cf.  Fig.~\ref{gs}).

Since stripes are vertical in cuprates for doping concentrations where
superconductivity occurs and diagonal in nickelates, we study both
orientations with arbitrary integer stripe spacing $p$.  Furthermore,
since charge order seems to be static up to high temperatures, in YBCO
up to 300K,\cite{Mook+02} holes can be considered as immobile at low
temperatures.  Our analysis is restricted to $T=0$.

We are interested in collective excitations around a ground state,
which -- for classical spins -- could be represented by
$\S(\br)=S\{0,0,\sigma(\br)\}$ with $\sigma=\pm 1$ on the electron
positions and $\sigma=0$ on the hole positions (as illustrated in
Fig.~\ref{gs}).  Denoting by $\bA^\1$ and $\bA^\2$ the primitive basis
vectors of the magnetic unit cell and by $\bA=m_1\bA^\1+m_2\bA^\2$ an
arbitrary magnetic lattice vector, the classical spin variables obey
the translational symmetry $\sigma(\br)=\sigma(\br+\bA)$. By placing
the origin at a hole position we obtain the additional reflection
symmetry $\sigma(\br)=-\sigma(-\br)$.

For a paradigmatic and minimalistic description of magnetic quantum
fluctuations we use a spin-only model with pair exchange.  More
complicated exchange processes such as cyclic
exchange\cite{Sugai+90,Muller-Hartmann+02} may be important for
quantitative purposes but are ignored here for simplicity.  We use a
generalized Heisenberg model on the two-dimensional square
lattice\cite{Tworzydlo+99}
\begin{equation}
  \mathcal H=\frac{1}{2}\sum'_{\br,\br'}J(\br,\br')\S(\br)\S(\br'),
\end{equation}
where the primed sums run over all spin positions with $\sigma \neq
0$. The exchange couplings obey the symmetry relations
\begin{subequations}
\begin{eqnarray}
  J(\br,\br') & = & J(\br',\br),\\
  J(\br,\br') & = & J(\br+\bA,\br'+\bA),\\
  J(\br,\br') & = & J(-\br,-\br').
\end{eqnarray}
\label{symm}
\end{subequations}
In fact, the exchange couplings may have a higher symmetry
corresponding to the hole lattice, which however will not be needed
explicitly in the further analysis.  To implement that the hole
strings act as antiphase boundaries between antiferromagnetic domains,
we assume that $J(\br,\br') = J>0$ for nearest neighbors $\br$, $\br'$
within the domains and $J(\br,\br') = \lambda J>0$ for nearest
neighbors across a string.

While it is natural to assume that $J$ should be comparable to the
exchange coupling in the undoped material, the coupling $\lambda J$
may deviate significantly.  To keep the number of parameters small, we
ignore that the exchange coupling even within an antiferromagnetic
domain should depend on the position of the pair relative to the hole
strings.

\section{Spin-wave Theory}
\label{sec.SWT}

We address the spin dynamics in the framework of linear spin-wave
theory (for a review in the context of cuprates, see e.g. Ref.
\onlinecite{Manousakis91}).  In the following analytic part we keep
the general form of the model and specialize to specific stripe
configurations later in Sec.~\ref{sec.res} when we numerically
evaluate the results of this section.  From now on all lengths are
expressed in units of the square-lattice spacing $a$.

\subsection{Holstein-Primakoff representation}

In the first step we flip all spins on one sublattice by
\begin{subequations}
\begin{eqnarray}
  S^x(\br) & = & \sigma^2(\br)\Sx(\br), \\
  S^y(\br) & = & \sigma(\br)\Sy(\br), \\
  S^z(\br) & = & \sigma(\br)\Sz(\br).
\end{eqnarray}
\end{subequations}
This transformation preserves the spin commutator relations.  Thereby,
we allow $\tilde \S$ to have spin $S$ also at the hole sites.
Although this introduces certain modes of zero energy, as we will
discuss below, it is advantageous to use a $\tilde \S$ with a
\textit{homogeneous} ferromagnetic ground state.

The corresponding transformed Hamiltonian reads
\begin{eqnarray}
  \mathcal H & = & \frac{1}{2}\sum_{\br,\br'}
  \tilde J(\br,\br')
  \left[\Sz(\br)\Sz(\br')+\Sy(\br)\Sy(\br')\right.\nn\\
  & & \phantom{\frac{1}{2}\sum_{\br,\br'}\tilde J(\br,\br')}
  \left. + \sigma(\br)\sigma(\br')\Sx(\br)\Sx(\br')\right],
\end{eqnarray}
where we have defined the new couplings $\tilde J(\br,\br') :=
J(\br,\br')\sigma(\br)\sigma(\br')$ which obey the same symmetry
relations (\ref{symm}) as $J$.

In the next step we represent the spin operators by the usual
Holstein-Primakoff (HP) bosons via
\begin{subequations}
\begin{eqnarray}
  \Sp & = & \sqrt{2S-\hat{n}}\phantom{.}b,\\
  \Sm & = & \bd\sqrt{2S-\hat{n}},\\
  \Sz & = & -\hat{n}+S,
\end{eqnarray}
\end{subequations}
with $\Spm=\Sx\pm i\Sy$. The eigenstates of the number operator
$\hat{n}=\bd b$ are restricted to $n\leq2S$ and the HP-operators
fulfill the canonical commutator relations $[b,\bd]=1$.  The
linearized spin-wave Hamiltonian $\cH$ is given by the terms quadratic
in the bosonic operators,
\begin{subequations}
\begin{eqnarray}
  \cH& = & \frac{S}{2}\sum_{\br,\br'}\left\{f(\br,\br')
    \left[\bd(\br)b(\br')+b(\br)\bd(\br')\right]\right.\nn
  \\
  & & +\left.g(\br,\br')\left[b(\br)b(\br')+\bd(\br)\bd(\br')\right]\right\},
  \\
  f(\br,\br') & = &\frac{1}{2} \tilde J(\br,\br')
  \left[\sigma(\br)\sigma(\br')+1\right], \nn
  \\
  & & -\delta_{\br,\br'}\sum_{\br'}\tilde J(\br,\br')
  \\
  g(\br,\br') & = &\frac{1}{2} \tilde J(\br,\br')
  \left[\sigma(\br)\sigma(\br')-1\right].
\end{eqnarray}
\end{subequations}
Obviously the functions $f$ and $g$ again satisfy the symmetry
relations (\ref{symm}).  

For further manipulations it is useful to decompose a vector
$\br=\bA+\ba$ on the square lattice into a vector
$\bA=m_1\bA^\1+m_2\bA^\2$ on the magnetic lattice and a decoration
vector $\ba$.  The number of vectors $\ba$ is denoted by $N$ (the area
of the magnetic unit cell).  In momentum space, the reciprocal
magnetic basis $\bQ^{(i)}$ defines the corresponding magnetic
Brillouin zone ($\mathcal{BZ}$).  Wave vectors $\bk$ can be uniquely
decomposed into $\bk=\bQ+\bq$ with $\bq\in\mathcal{BZ}$ and
$\bQ=m_1\bQ^\1+m_2\bQ^\2$.  Within the Brillouin zone of the square
lattice there are $N$ vectors $\bQ$ which we denote by $\bQ_\nu$.

We Fourier transform the bosonic operators via $b(\br)=\int_\bk
\exp(i\bk\br)b(\bk)$, where $\int_\bk = (2 \pi)^{-2} \int d^2 k$
and the $\bk$ integrals run over the Brillouin zone of the square
lattice with an area $(2\pi)^2$.  
Using these decompositions and the Poisson sum formula
\begin{eqnarray}
\sum_{\bA} e^{i\bk\bA} & = & \frac{1}{N}\sum_{\bQ}\delta(\bk+\bQ)
\end{eqnarray}
we rewrite the spin-wave Hamiltonian as
\begin{eqnarray}
  \cH & = & \frac{1}{2}\int_\bq\sum_{\nu,\nu'}F_{\nu,\nu'}(\bq)
  [\bd_{\bq+\bQ_\nu}b_{\bq+\bQ_{\nu'}}\nn\\
  & & \qquad+b_{-\bq-\bQ_\nu}\bd_{-\bq-\bQ_{\nu'}}]\nn \\
  &  &+ \frac{1}{2}\int_\bq\sum_{\nu,\nu'}G_{\nu,\nu'}(\bq)
  [\bd_{\bq+\bQ_\nu}\bd_{-\bq-\bQ_{\nu'}}\nn\\
  & & \qquad+b_{-\bq-\bQ_\nu}b_{\bq+\bQ_{\nu'}}] ,
\end{eqnarray}
where
\begin{eqnarray}
  F_{\nu,\nu'}(\bq) & = & \frac{S}{N}\sum_{\bA}\sum_{\ba,\ba'}f(\ba+\bA,\ba')
  \nn\\
  & & \times \cos\left[\bq\bA+\bq(\ba-\ba')+\bQ_\nu\ba-\bQ_{\nu'}\ba'\right]
\end{eqnarray}
is essentially the Fourier transform of $f$, 
\begin{eqnarray}
  \frac{S}{N} f(\bQ_\nu+\bq,\bQ_{\nu'} + \bq') = 
  \delta(\bq+\bq')  F_{\nu,\nu'}(\bq) .
\end{eqnarray}
Analogous expressions relate $G$ to $g$.

\subsection{Bogoliubov Transformation}
\label{bogo}

To diagonalize the Hamiltonian, we express the bosonic operators by
canonical coordinate and momentum operators $\Phi_\nu(\bq) :=
\Phi(\bq+\bQ_\nu)$ and $\Pi_\nu(\bq) := \Pi(\bq+\bQ_\nu)$ via the
relations
\begin{subequations}
\begin{eqnarray}
  \Phi_\nu(\bq) & = & \frac 1{\sqrt2}
  \left(b_{\bq+\bQ_\nu}+\bd_{-\bq-\bQ_\nu}\right),\\
  \Pi_\nu(\bq) & = &  \frac 1{\sqrt2 i} 
  \left(b_{-\bq-\bQ_\nu} - \bd_{\bq+\bQ_\nu}\right) .
\end{eqnarray}
\end{subequations}
In terms of these operators, the spin-wave Hamiltonian reads
\begin{eqnarray}
  \cH &=& \frac 12 \int_\bq \sum_{\nu,\nu'} 
  \bigg\{ \Pid_\nu(\bq) {M^{-1}}_{\nu,\nu'}(\bq) \Pi_{\nu'}(\bq) 
  \nonumber \\ &&
  + \Phid_\nu(\bq) K_{\nu,\nu'}(\bq) \Phi_{\nu'}(\bq) \bigg\},
\end{eqnarray}
with the inverse mass matrix $\bM^{-1}=\bF-\bG$ and the coupling
matrix $\bK=\bF+\bG$. As a result of the invariance of the Hamiltonian
under the replacement $\Sx(\br)\to\sigma(\br)\Sx(\br)$,
$\Sy(\br)\to\sigma(\br)\Sy(\br)$ one can easily derive the symmetry
conditions
\begin{subequations}
\begin{eqnarray}
  \bK &=& \bsigma  \bM^{-1} \bsigma,\\
  \bM^{-1} &=& \bsigma  \bK \bsigma,
\end{eqnarray}
\end{subequations}
where we have introduced the hermitian matrix
$\sigma_{\nu,\nu'}:=\frac 1N \sum_\ba e^{-i(\bQ_\nu-\bQ_{\nu'})\ba}
\sigma(\ba)$. To simplify notation, we suppress arguments $\bq$ which
may be considered as fixed during the diagonalization in $\nu$ space
and use the pseudo Dirac notation $|\Phi\rrangle:=\sum_\nu
\Phi_\nu|\nu\rrangle$, $|\Pi\rrangle:=\sum_\nu \Pi_\nu|\nu\rrangle$
with the Cartesian basis $|\nu\rrangle$, $\nu=1,\ldots,N$.  After
performing the canonical transformation $|\Phi \rrangle=\bM^{-1/2}
|\tilde\Phi\rrangle$, $|\Pi \rrangle=\bM^{1/2} |\tilde\Pi\rrangle$ the
Hamiltonian can be rewritten as
\begin{eqnarray}
  \cH &=& \frac 12 \int_\bq
  \bigg\{ \llangle \tilde \Pi | \tilde \Pi \rrangle
  + \llangle \tilde\Phi | \bM^{-1/2} \bK \bM^{-1/2} | 
  \tilde \Phi \rrangle \bigg\},\qquad
\end{eqnarray}
and we still have to diagonalize $\bM^{-1/2} \bK \bM^{-1/2}=\bW^2$
with hermitian $\bW:=\bM^{-1/2} \bsigma \bM^{-1/2}$.  Introducing an
orthonormal eigenbasis $\{|\alpha\rrangle,\alpha=1,\ldots N\}$ of this
matrix, $\bW|\alpha\rrangle=\xi_\alpha|\alpha\rrangle$, and defining
$\w_\alpha := |\xi_\alpha|$ we can transform to normal coordinates
\begin{subequations}
\begin{eqnarray}
  \tilde \Phi_\nu &=& \sum_\alpha \w_\alpha^{-1/2} 
  \llangle \nu|\alpha\rrangle \tilde \Phi_\alpha,
  \\
  \tilde \Pi_\nu &=& \sum_\alpha \w_\alpha^{1/2} 
  \llangle \alpha|\nu\rrangle \tilde \Pi_\alpha,
\end{eqnarray}
\end{subequations}
and obtain
\begin{eqnarray}
 \cH = \frac 12  \sum_\alpha \int_\bq \w_\alpha \left\{
    \tPid_\alpha \tilde \Pi_\alpha 
    + \tilde \Phi^\dagger_\alpha \tilde \Phi_\alpha \right\}.
\end{eqnarray}
Transforming back to corresponding bosonic operators
$\tilde\Phi_\alpha(\bq)=\frac 1{\sqrt2} [b_\alpha(\bq) +
\bd_\alpha(-\bq)]$, $\tilde\Pi_\alpha(\bq)= \frac 1{\sqrt2 i}
[b_\alpha(-\bq) - \bd_\alpha(\bq)]$ we obtain the final diagonal
bosonic representation of the spin-wave Hamiltonian
\begin{eqnarray}
  \cH = \sum_{\alpha} \int_\bq \omega_\alpha(\bq) 
  \bigg\{\frac 12 + {\bd_\alpha}(\bq)  b_\alpha(\bq) \bigg\}.
\end{eqnarray}
Thus, as the result of the above diagonalization we obtain
$\w_\alpha(\bq)$ as the magnon dispersion relation with the band index
$\alpha$.

We would like to remark that the $|\nu\rrangle$ space contains a
common subspace of eigenvectors of the matrices $\bsigma$, $\bM^{-1}$
and $\bK$ with vanishing eigenvalues.  This subspace is $h$
dimensional, where $h$ is the number of holes in the magnetic unit
cell.  These zero modes are an artifact of the introduction of spins
$\tilde \S$ on the hole sites.  All above manipulations, including
e.g. the calculation of $\bM^{1/2}$ and $\w_\alpha^{-1}$, are
welldefined on the orthogonal subspace of physical spins.

\subsection{Structure Factor}

In this section we proceed to calculate the zero-temperature structure
factor
\begin{eqnarray}
  \cS(\bk,\w) &:=& \sum_\fin \sum_{j=x,y,z}  
  |\langle \fin | S^j(\bk)|0 \rangle |^2 \delta (\w-\w_\fin) .
\end{eqnarray}
Here, $|0\rangle$ denotes the ground state (magnon vacuum)
characterized by $b_\alpha(\bq) |0\rangle =0$ and we consider only
single-magnon states $|\fin\rangle$ with excitation energy
$\w_\fin:=E_\fin-E_0$.  Since
\begin{eqnarray}
  S^z(\bk) &=& S \sum_{\nu'} \delta(\bk-\bQ_{\nu'}) \sigma(\bQ_{\nu'}) 
  \nn \\ &&
  - \sum_{\nu'} \sigma(\bQ_{\nu'}) \int_{\bk''} 
  \bd(\bk'') b(\bk-\bQ_{\nu'}+\bk'')\qquad
\end{eqnarray}
with $\sigma(\bQ):=\frac 1N \sum_\ba e^{-i \bQ \ba} \sigma(\ba)$
preserves the magnon number, it contributes only to the elastic part
of the structure factor,
\begin{eqnarray}
  \cS^\textrm{el}(\bk,\w) \propto S^2\sum_\bQ 
  \delta(\bk-\bQ) |\sigma(\bQ)|^2 \delta(\w).
\end{eqnarray}
To calculate the inelastic part of the structure factor (which has
contributions of order $S$ only from $j=x,y$) we express these spin
components by the bosonic operators using the transformations derived
in section~\ref{bogo},
\begin{subequations}
\begin{eqnarray}
  S^x(\bq+\bQ_\nu) &\approx& \sqrt S \sum_{\nu'}  
  \sigma^\2(\bQ_\nu-\bQ_{\nu'})
  \Phi_{\nu'}(\bq)\nn\\
  & = & \sqrt{\frac S2}\sum_{\alpha,\nu'}\sigma^\2(\bQ_\nu-\bQ_{\nu'})
  \w_\alpha^{-1/2} \nn\\
  & & \times \llangle \nu' | \bM^{-1/2} | \alpha 
  \rrangle [b_\alpha(\bq)+\bd_\alpha(-\bq)],\qquad\\
  S^y(\bq+\bQ_\nu) &\approx& \sqrt S \sum_{\nu'}  \sigma(\bQ_\nu-\bQ_{\nu'})
  \Pid_{\nu'}(\bq)\nn\\
  & = & i\sqrt{\frac S2}\sum_{\alpha,\nu'}\sigma(\bQ_\nu-\bQ_{\nu'})
  \w_\alpha^{1/2}\nn\\
  & & \times\llangle \nu'| \bM^{1/2} | \alpha 
  \rrangle [b_\alpha(\bq)-\bd_\alpha(-\bq)],\qquad
\end{eqnarray}
\end{subequations}
where we have defined $\sigma^\2(\bQ):=\frac 1N \sum_\ba e^{-i \bQ
  \ba} \sigma^2(\ba)$. Since the contributing final states are just
given by the one-magnon states $|\fin \rangle=\bd_\alpha(\bq)|0
\rangle$ it is easy to calculate the inelastic part of the structure
factor. Using the relations $\bsigma^2 \bM^{-1/2} |\alpha\rrangle =
\bsigma \bM^{1/2} \bW|\alpha\rrangle = \xi_\alpha \bsigma \bM^{1/2}
|\alpha\rrangle$ and $\bsigma^2 \bM^{-1/2} |\alpha \rrangle =
\bM^{-1/2} |\alpha \rrangle$ we obtain
\begin{subequations}
\begin{eqnarray}
  \cS^\textrm{in}(\bq+\bQ_\nu,\w) &=&  S \sum_{\alpha} 
  S_\alpha(\bq+\bQ_\nu)  \delta (\w-\w_\alpha(\bq)),\qquad
  \\
  S_\alpha(\bq+\bQ_\nu)&=&\llangle \nu| \bM^{-1/2} |\alpha \rrangle
  \frac 1{\w_\alpha} \llangle \alpha |\bM^{-1/2} |\nu \rrangle.\qquad 
\end{eqnarray}
\label{strucf}
\end{subequations}
At this point it may be helpful to remind that $\bq$ is an implicit
argument of $\w_\alpha$, $\bM^{-1/2}$ and $|\alpha\rrangle$.  The
periodicity $\w_\alpha(\bq) = \w_\alpha(\bq+\bQ)$ of the
eigenfrequencies is absent in the structure factor since the coupling
of an external field to a spin wave wave vector $\bk= \bq+\bQ$ depends
on $\bQ$.

\section{Results}
\label{sec.res}

We now evaluate the above general analytic expressions for the magnon
dispersion and the structure factor.  Thereby we focus on our
minimalistic model (cf. Sec.~\ref{sec.model}) with stripe spacings
$p=3$, $4$, and $5$, since these values correspond to doping
concentrations in various experimental works as mentioned in the
introduction.  The explicit comparison to experiments is postponed to
Sec.~\ref{sec.disc}.

For later reference, we briefly recall that for the undoped
two-dimensional antiferromagnet (which is recovered by our model in
the limit $p\to\infty$) the spin-wave dispersion is given by
\begin{equation}
  \w_{\textrm{AF}}(\bk)=2JS
  \sqrt{4-\left[\cos(2\pi H)+\cos(2\pi K)\right]^2}.
\end{equation}
[From now on, we refer to wave vectors $\bk=(H,K)$ in units of
$2\pi/a$].  It vanishes at the antiferromagnetic wave vector
$\bk_{\textrm{AF}}=(\frac12,\frac12)$, where the structure factor
shows maximal intensity To leading order in $\delta
\bq=\bk-\bk_{\textrm{AF}}$, the low energy spin-wave excitations are
characterized by an isotropic dispersion $\w_{\textrm{AF}}\approx
v_{\textrm{AF}}|\delta \bq|$ with a spin-wave velocity
$v_{\textrm{AF}}= \sqrt8 JSa$.

\subsection{Vertical case}

For vertical stripes a possible magnetic unit cell is given by the
basis vectors $\bA^\1=(0,2)$ and $\bA^\2=(p,0)$ for odd or
$\bA^\2=(p,1)$ for even $p$. Therefore we have $N=2p$ lattice sites
per unit cell (cf. Fig.~\ref{gs}) and $2p$ eigenvalues
$\w_\alpha(\bq)$.  Two of them (corresponding to the number of holes)
vanish identically and we obtain $p-1$ twofold degenerate physical
bands. This degeneracy results from the equivalence of the two
sublattices.

\begin{figure}[h]
\epsfig{figure=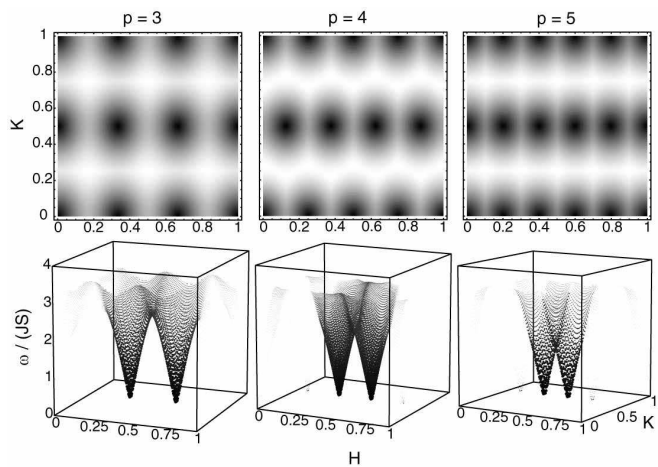,width=\linewidth}
\caption{Acoustical band for vertical stripes with spacings $p=3,4,5$ and
  $\lambda=0.5$.  The upper row shows density plots of the spin-wave
  dispersion, where dark regions correspond to low energy values.  The
  lower row shows the acoustical band in the $(H,K,\w)$ space
  including the weight of the inelastic structure factor, where larger
  weight corresponds to darker points with larger size.}
\label{fig.v_HK}
\end{figure}

The lowest, acoustical band has zeros at the magnetic superstructure
lattice vectors.  Within the Brillouin zone of the square lattice (we
choose $0 \leq H,K < 1$), the vectors $\bQ_\nu$ are located at $(\frac
jp,0)$ and $(\frac jp, \frac 12)$ for odd $p$ or $(\frac jp +\frac
1{2p},\frac 12)$ for even $p$ (with $0 \leq j <p$).  In the upper row
of Fig.~\ref{fig.v_HK}, we show $\w(\bk)$ for the acoustical band as a
density plot, where black corresponds to $\w=0$ and white to the upper
band edge.

Although the dispersion relation obeys the symmetry
$\w(\bk)=\w(\bk+\bQ)$ corresponding to the period of the magnetic unit
cell, this symmetry is absent in the structure factor.  In the lower
row of Fig.~\ref{fig.v_HK} the acoustic band is replotted in the
$(H,K,\w)$ space using darker and thicker dots for points with larger
values of the structure factor (\ref{strucf}).  In agreement with
experiments, the weights are concentrated near the lowest harmonic
``incommensurate'' wave vectors $\bQ=(\frac 12 \pm \frac 1{2p}, \frac
12)$.  Higher harmonics of the superstructure are much weaker as
already noticed in Ref.~\onlinecite{Tranquada+95}.   

\begin{figure}[h]
\epsfig{figure=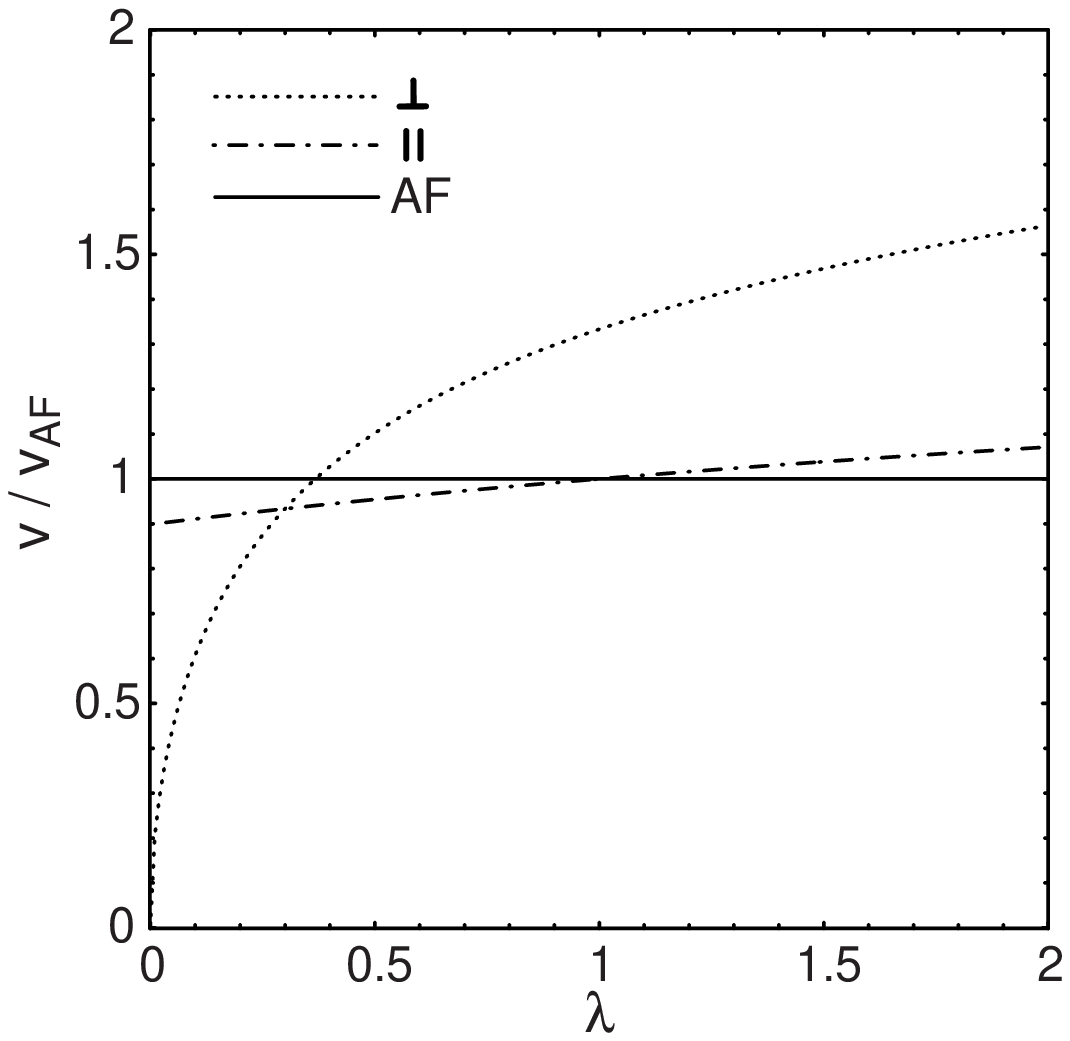,width=0.46\linewidth}
\hspace{0.01\linewidth}
\epsfig{figure=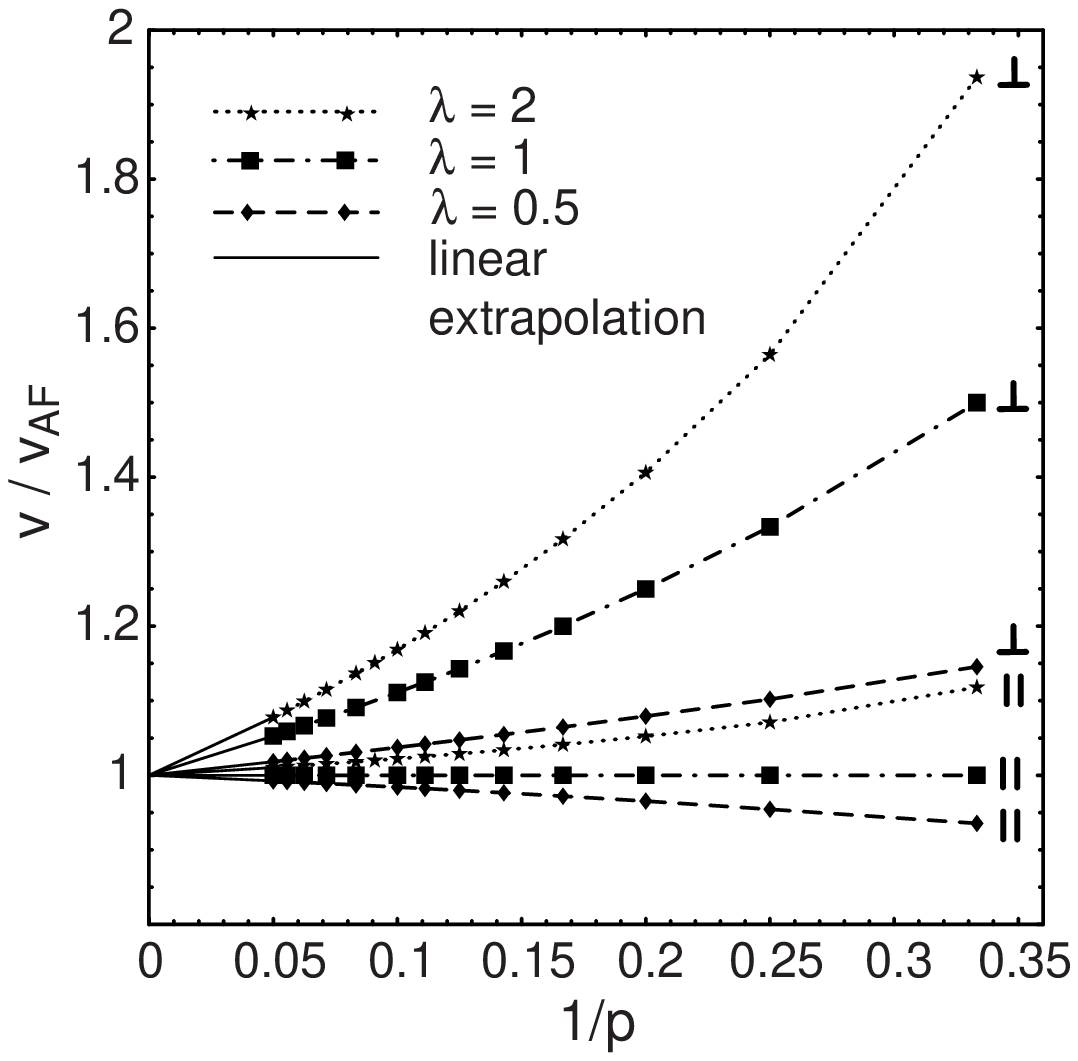,width=0.46\linewidth}
\caption{Spin-wave velocities $v_\perp$ and $v_\parallel$ for vertical 
  stripes with spacing $p=4$ as a function of $\lambda$ (left panel)
  and as a function of $1/p$ for different couplings $\lambda$ (right
  panel; lines are a guide to the eye).}
\label{fig.v_v}
\end{figure}

To study the anisotropy of the dispersion next to the satellite
positions we calculate the spin-wave velocities $v_\perp$ and
$v_\parallel$ perpendicular and parallel to the stripe orientation
(cf. Fig.~\ref{fig.v_v}).  For $\lambda=0$, where the coupling between
the domains is switched off, $v_\perp$ is zero and $v_\parallel$
remains finite. With increasing $\lambda$ both velocities increase,
$v_\perp$ more strongly than $v_\parallel$. There exists a value
$\lambda^*$ with isotropic velocities, $v_\perp=v_\parallel$. For
$p=4$ we find $\lambda^*\approx 0.3$. in the limit $p\to\infty$ both
velocities converge to $v_{\textrm{AF}}$ as expected, for $p \gg 1$ we
find $v_{\perp,\parallel}/v_{\textrm{AF}}-1 \propto 1/p$.  In the
special case $\lambda=1$ the velocities are given by
\begin{subequations}
\begin{eqnarray}
  v_\parallel & = & v_{\textrm{AF}} ,\\
  v_\perp & = &  \frac{p}{p-1}v_{\textrm{AF}}
\end{eqnarray}
\label{v}
\end{subequations}
for purely geometric reasons.  In this case, all spins are interacting
in terms of the topology and strength of the exchange couplings
exactly like in the antiferromagnet. The only difference lies in the
insertion of strings of holes, which effectively stretch the lattice
and increase the velocity by a factor $p/(p-1)$ in the perpendicular
direction.

\begin{figure}[h]
\epsfig{figure=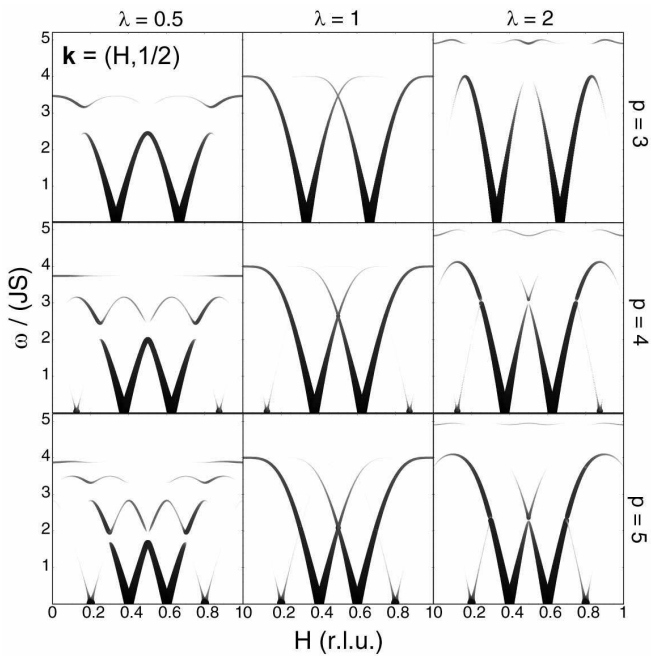,width=\linewidth}
\caption{Band structure for vertical stripes along the $(H,0.5)$ direction
  with different spacings $p$ and couplings $\lambda$. Darker and
  larger points correspond to a larger weight of the inelastic
  structure factor.}
\label{fig.v_H}
\end{figure}

We now focus on the line $\bk=(H,\frac12)$ containing the satellites,
along which we plot all $p-1$ magnon bands in Fig.~\ref{fig.v_H} for a
variety of $p$ and $\lambda$.  For $\lambda<1$ and $\lambda>1$ the
bands are separated by gaps. (In this context, ``gaps'' are not
necessarily real gaps showing up in the density of states, they are
apparent gaps along the chosen line.)  Only for $\lambda=1$, the
structure seems to consist of displaced and intersecting
antiferromagnetic bands.  The value $\lambda=1$ is special for the
reasons explained above which also imply that the band width must
coincide with the antiferromagnet.  The purely geometric effect
entails just a more complicated band structure.

\begin{figure}[h]
\epsfig{figure=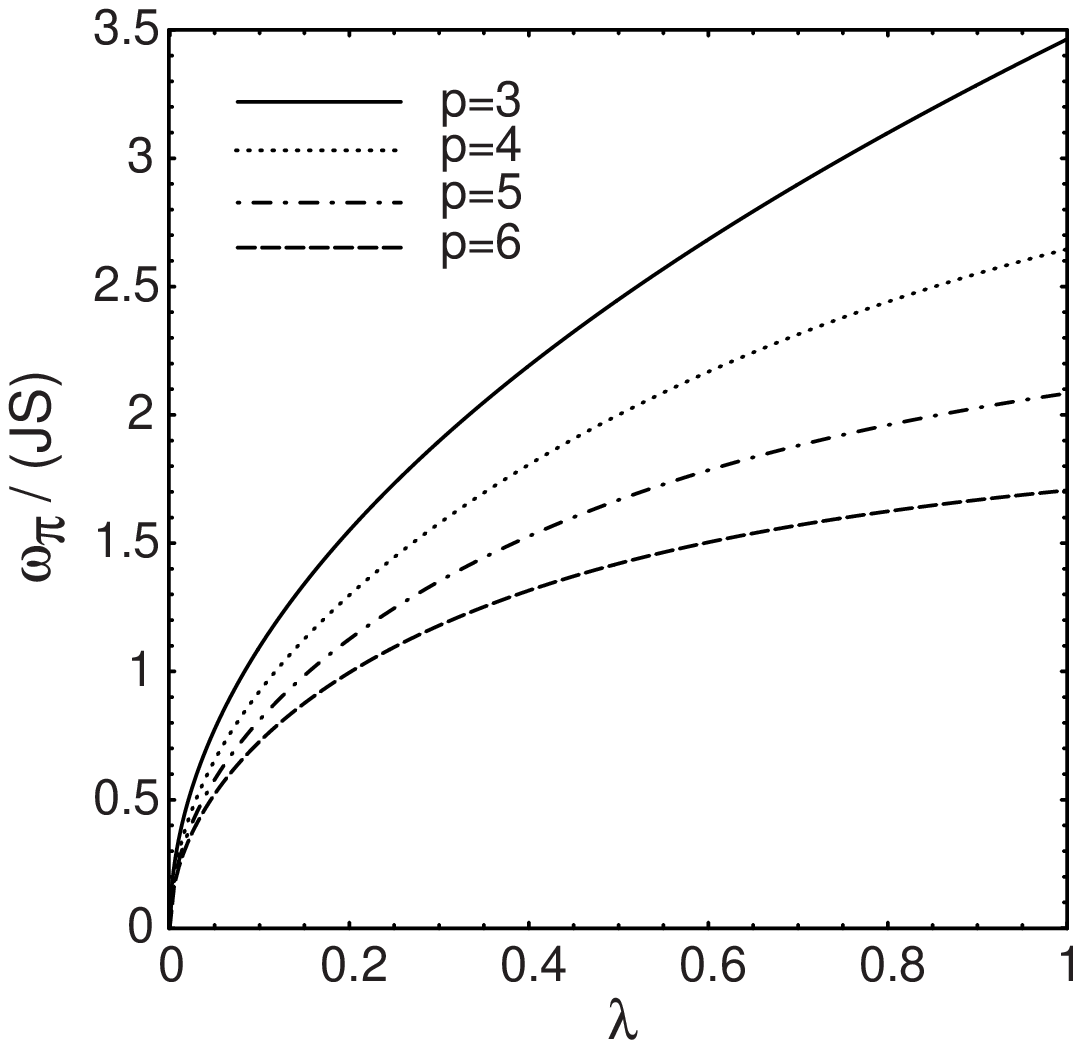,width=0.46\linewidth}
\hspace{0.01\linewidth}
\epsfig{figure=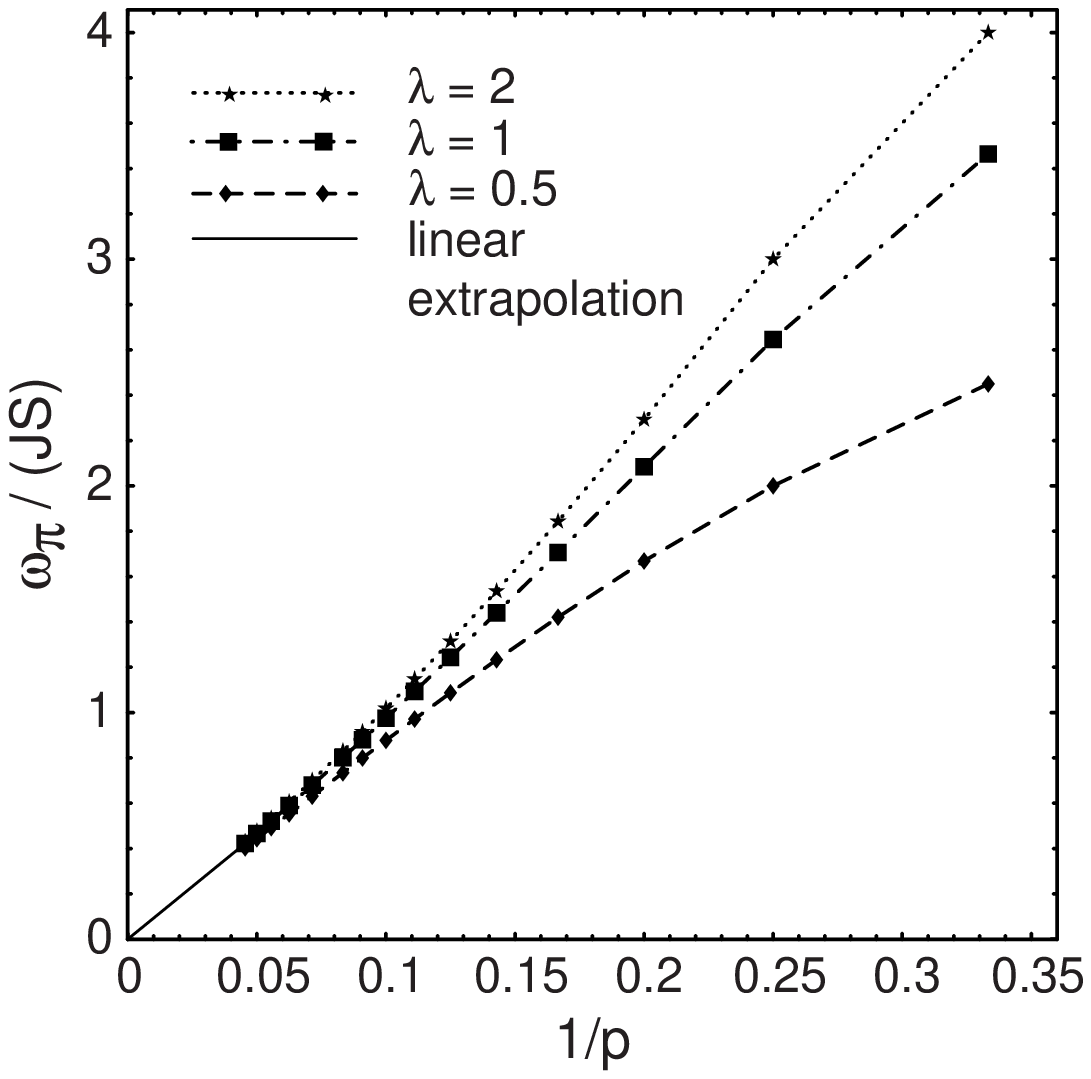,width=0.46\linewidth}
\caption{The resonance frequency $\w_\pi$ for vertical stripes as a 
  function of $\lambda$ for different spacings $p$ (left) and as a
  function of $1/p$ for different couplings $\lambda$ (right; lines
  are a guide to the eye).}
\label{fig.v_w}
\end{figure}

To the extent to which our stripe model provides a valid description
of the magnetic excitations in the materials where the $\pi$ resonance
was observed, the resonance frequency has to be identified with
$\w(\bk_\AF)$ from the lowest magnon band, provided $\w(\bk_\AF)>0$
and the structure factor has significant weight.  From
Fig.~\ref{fig.v_H} one recognizes that for $\lambda <1$ this is always
the acoustical band.  On the other hand, for $\lambda>1$ higher bands
may yield a stronger resonance (see case $p=3$ and $\lambda=2$).  In
Fig.~\ref{fig.v_w} we illustrate the dependence of $\w_\pi$ on
$\lambda$ and $p$.

For $p$ large enough such that $v_\parallel \approx v_\AF$ and the
magnon dispersion is roughly linear between the main satellite and
$\bk_\AF$, we may estimate
\begin{eqnarray}
  \w_\pi \approx v_\AF \frac \pi{pa}.
\end{eqnarray}
This estimate becomes exact for small $1/p$ and represents the linear
asymptotics in Fig.~\ref{fig.v_w} (right).  Deviations grow with
decreasing $p$ and increasing deviation of $\lambda$ from 1.

\subsection{Diagonal case}

For diagonal stripes there are more subtle differences between even
and odd stripe spacings $p$.  Since the basis vectors of the magnetic
unit cell can be chosen as $\bA^\1=(-1,1)$ and $\bA^\2=(p,0)$ for odd
or $\bA^\2=(2p,0)$ for even $p$ (cf. Fig.~\ref{gs}), we have one hole
and $p-1$ spins per unit cell for odd $p$ and twice the number of
holes and spins for even $p$.  Like in the vertical case, the number
of eigenvalues vanishing identically corresponds to the number of
holes, the number of bands is given by half of the number of spins per
unit cell, and the bands are twofold degenerate.

\begin{figure}[h]
  \epsfig{figure=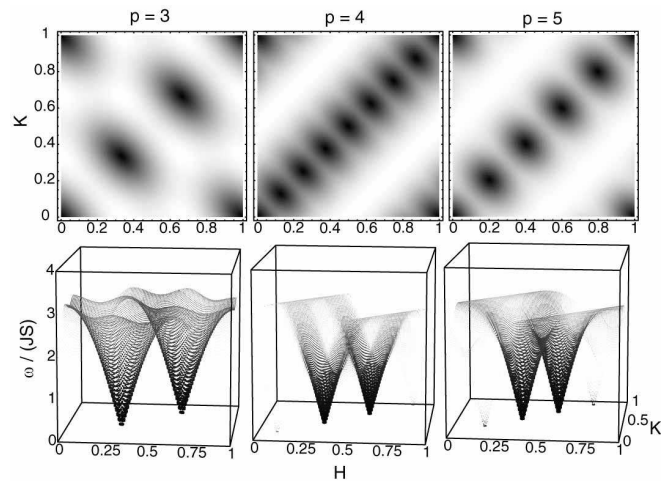,width=\linewidth}
\caption{Acoustical band for diagonal stripes with spacings $p=3,4,5$ for 
  $\lambda=1$ plotted in analogy to Fig.~\ref{fig.v_HK}. }
\label{fig.d_HK}
\end{figure}

All magnetic Bragg peaks are located along the line $\bQ=(H,H)$ with
$H=j/p$ for odd and $H=j/(2p)$ for even $p$ (cf.
Fig.~\ref{fig.d_HK}). In the case $p=3$ we can calculate the
dispersion analytically and find
\begin{eqnarray}
  \w(\bk) & = & 2JS\big\{\sin^2\left[\pi(H-K)\right]+
  \lambda\sin^2\left[\pi(2H+K)\right]\nn \\
  & & +\lambda\sin^2\left[\pi(H+2K)\right]\big\}^{1/2}
  .\qquad
\end{eqnarray} 
Along the $\bk=(H,H)$ direction this relation simplifies to
\begin{equation}
  \w(H,H)= \sqrt{2\lambda} JS |\sin(3\pi H)|.
  \label{sinu}
\end{equation}

\begin{figure}[h]
\epsfig{figure=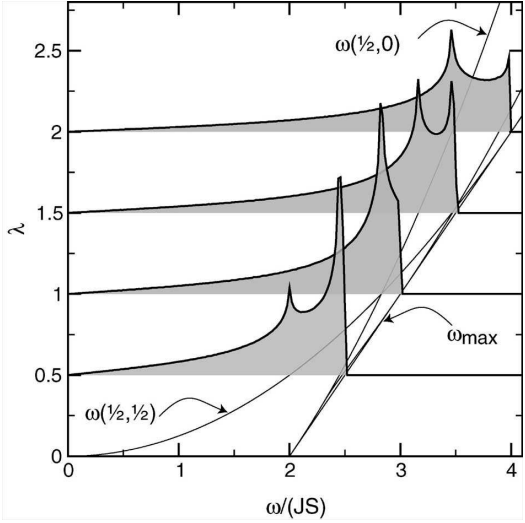,width=0.8\linewidth}
\caption{Density of states $\rho(\w)$ 
  (height of shaded area in arbitrary units) for diagonal stripes with
  spacing $p=3$ and different couplings $\lambda$. The thin lines
  correspond to the energies of the critical points in the dispersion.
  There are up to four inequivalent ones, at $\bk=(\frac 12,\frac12)$,
  $\bk=(\frac 12,0)$, the upper band edge with $\w_\textrm{max}$ and a
  possible additional critical point.}
\label{fig.zust}
\end{figure}

Though the case $p=3$ with a single band is the simplest possible, we
find several critical points in the dispersion, which should result in
a nontrivial shape of the density of states $\rho(\w) \sim
\int_{\bk}\delta(\w-\w(\bk))$. Therefore we calculate this quantity
just to illustrate that even for this simplest case $\rho(\w)$ shows
interesting features strongly depending on the effective coupling
$\lambda$.  The numerically calculated density of states is plotted in
Fig.~\ref{fig.zust} for different values of $\lambda$. The van-Hove
singularities are located at the energies of the critical points in
the dispersion.  The dependence of these energies on the coupling
$\lambda$ is also shown in this figure.  Due to a finite numerical
resolution the van-Hove singularities are not resolved if they are too
close to each other and their precise shape is not reproduced, e.g. at
the energies of the saddle points $\rho(\w)$ should diverge
logarithmically.
 
\begin{figure}[h]
\epsfig{figure=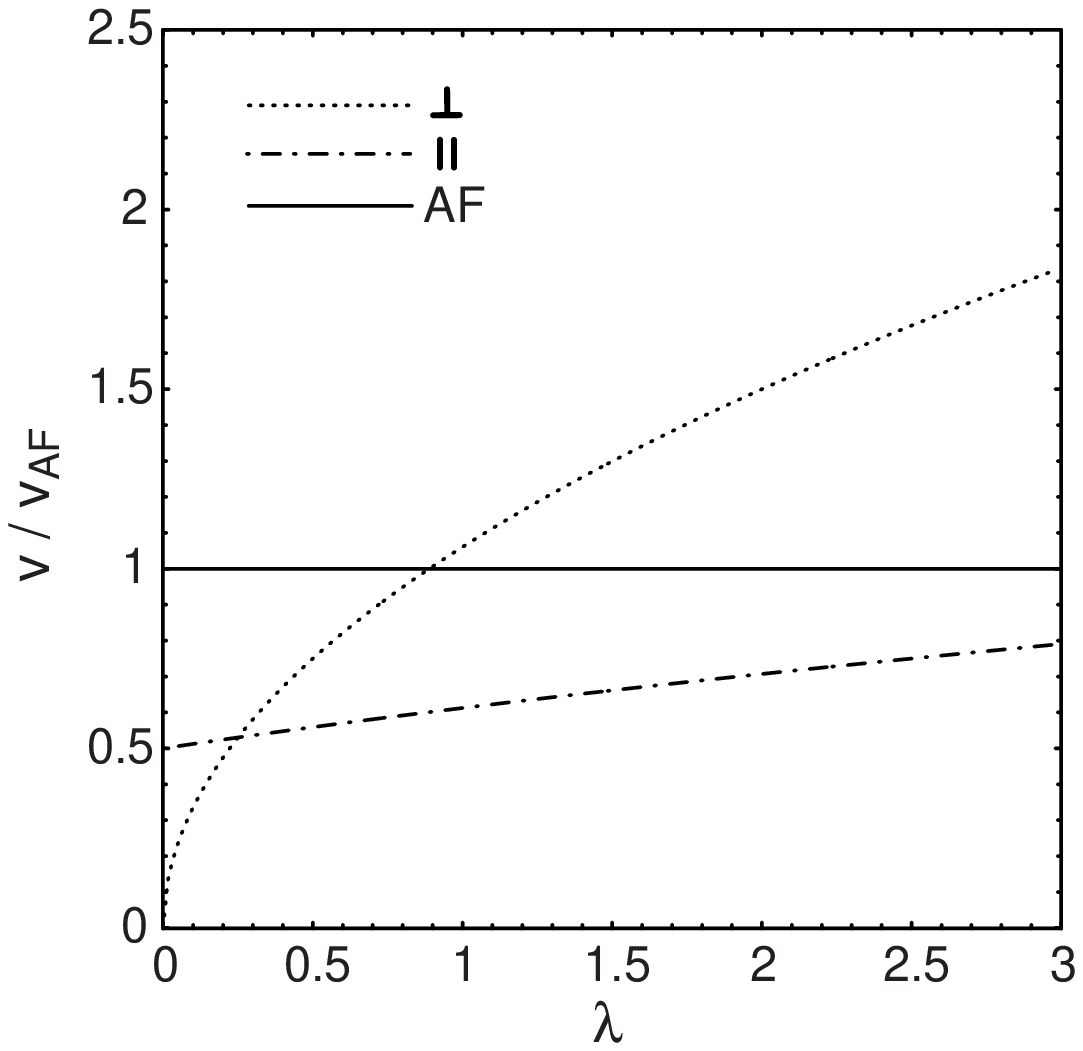,width=0.46\linewidth}
\hspace{0.01\linewidth}
\epsfig{figure=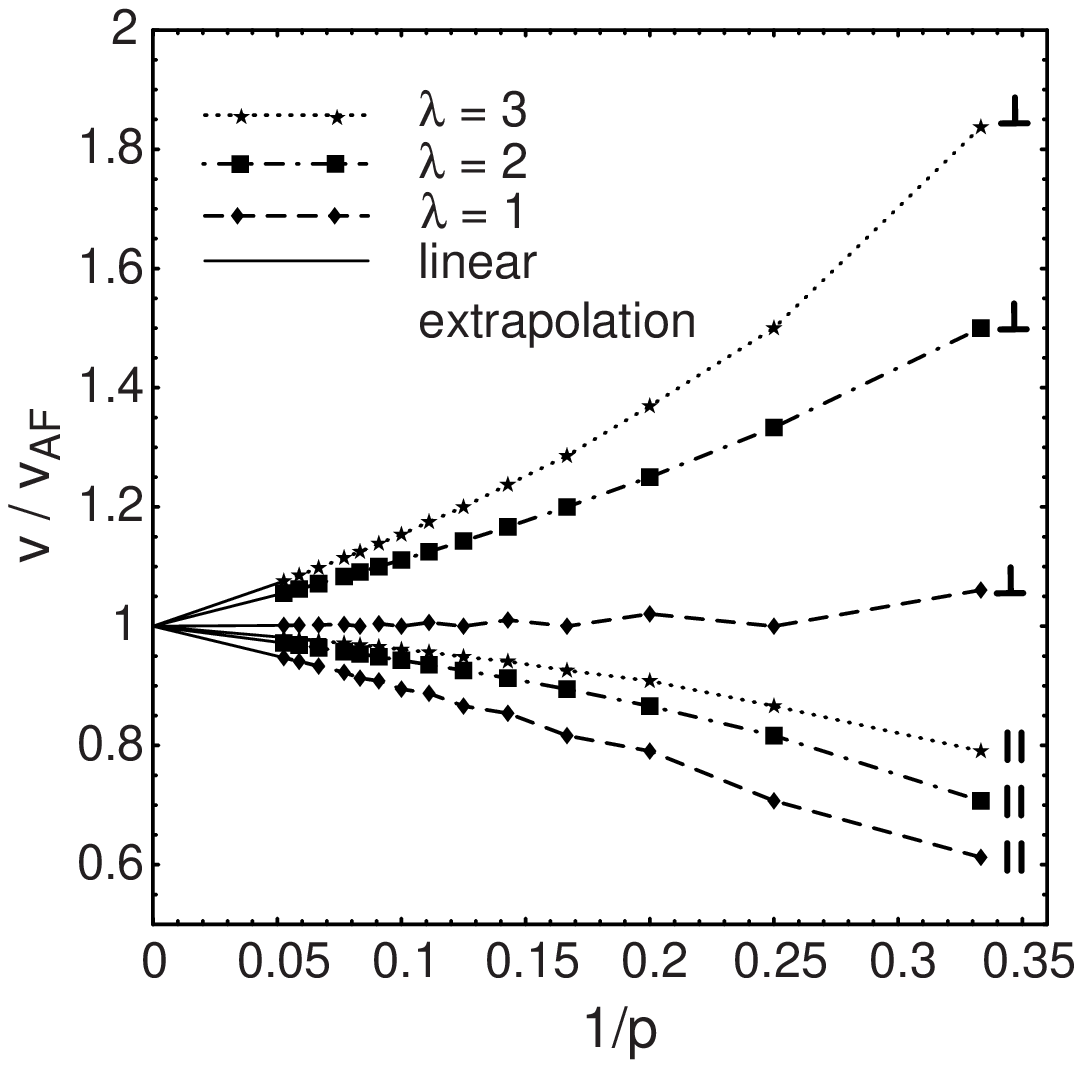,width=0.46\linewidth}
\caption{Spin-wave velocities $v_\perp$ and $v_\parallel$ for 
  diagonal stripes with spacing $p=3$ as a function of $\lambda$
  (left) and as a function of $1/p$ for different couplings $\lambda$
  (right; lines are a guide to the eye).}
\label{fig.d_v}
\end{figure}

Calculating the weight by the structure factor of the bands we find
the strongest intensity near the zeros of the acoustic band at the
satellite positions $\bQ=(\frac 12 \pm \frac 1{2p} ,\frac 12 \pm \frac
1{2p})$ for all $p$.  The behavior of the spin-wave velocities
$v_\perp$ and $v_\parallel$ (cf.  Fig.~\ref{fig.d_v}) is similar to
the vertical case.

Along the $\bk=(H,H)$ direction, the acoustical band is separated by
finite gaps from the optical bands for $\lambda\neq 2$. For
$\lambda=2$, the band structure again seems to consist of intersecting
displaced antiferromagnetic bands.  In contrast to the vertical case,
the special value of $\lambda$ is now $2$ since for this value the sum
of the exchange couplings to neighboring spins is as large as in the
antiferromagnet.  However, for diagonal stripes the topology of the
couplings is different from the antiferromagnet.

\begin{figure}[h]
\epsfig{figure=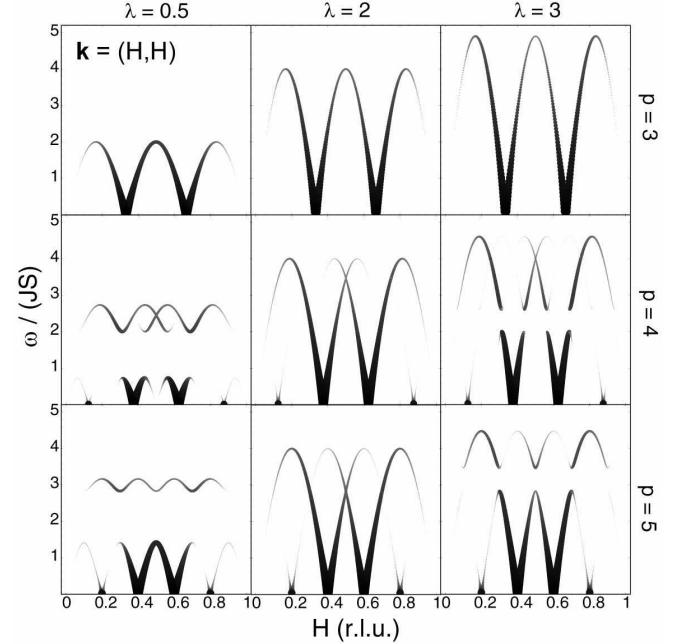,width=\linewidth}
\caption{Band structure for diagonal stripes along the $(H,H)$ direction
  with different spacings $p$ and couplings $\lambda$.}
\label{fig.d_HH}
\end{figure}

For odd $p$, the $\pi$ resonance results from the excitation of
acoustical magnons since the lowest band has a finite
$\w(\bk_{\textrm{AF}})$ with a relatively strong weight.  In contrast,
for even $p$ the frequency and the weight of the acoustical band
vanish at $\bk_{\textrm{AF}}$.  In this case, the $\pi$ resonance
should therefore be ascribed to optical magnons.  For $\lambda=2$, the
$\pi$ resonance results from the common edge of the acoustical and
optical bands (cf.  Fig.~\ref{fig.d_HH}).

\begin{figure}[h]
\epsfig{figure=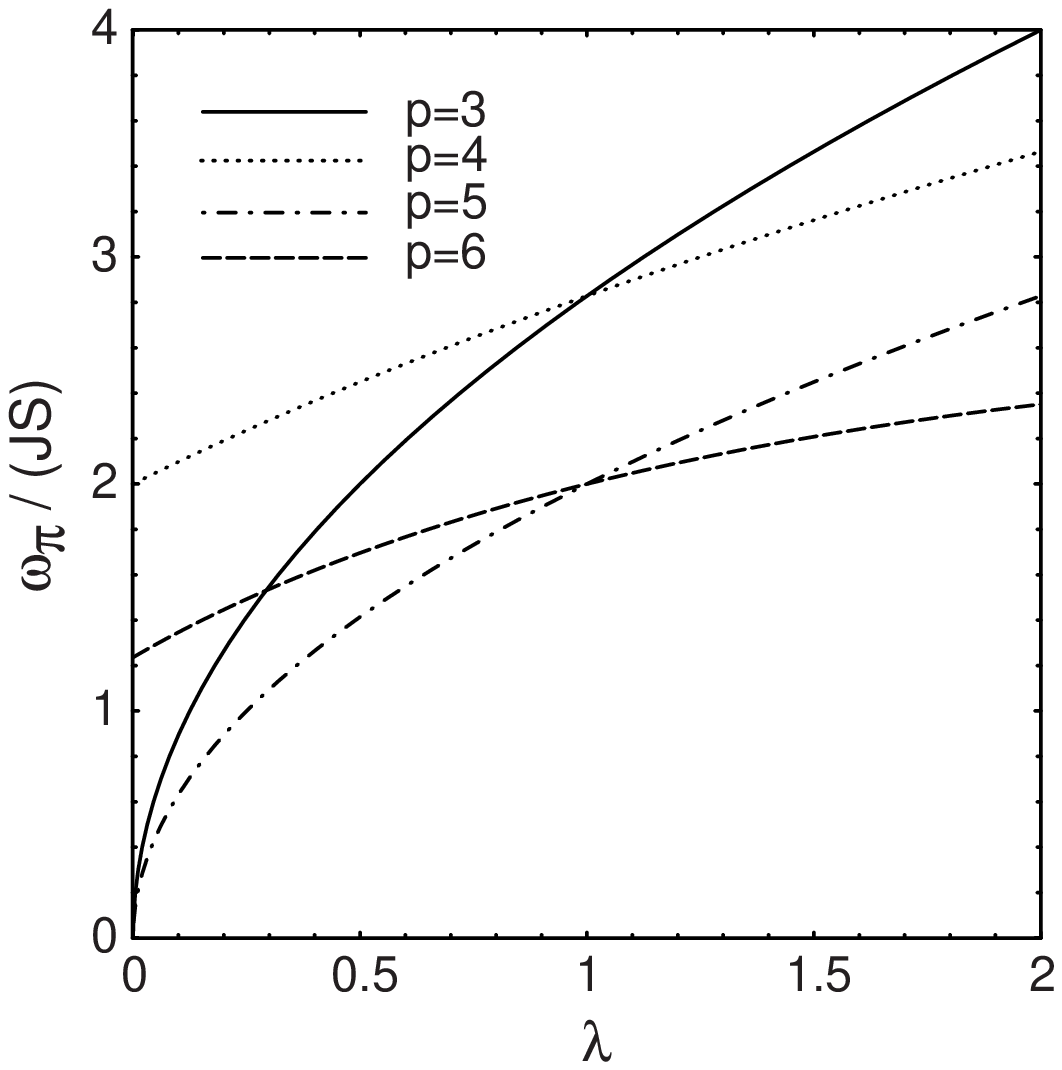,width=0.46\linewidth}
\hspace{0.01\linewidth}
\epsfig{figure=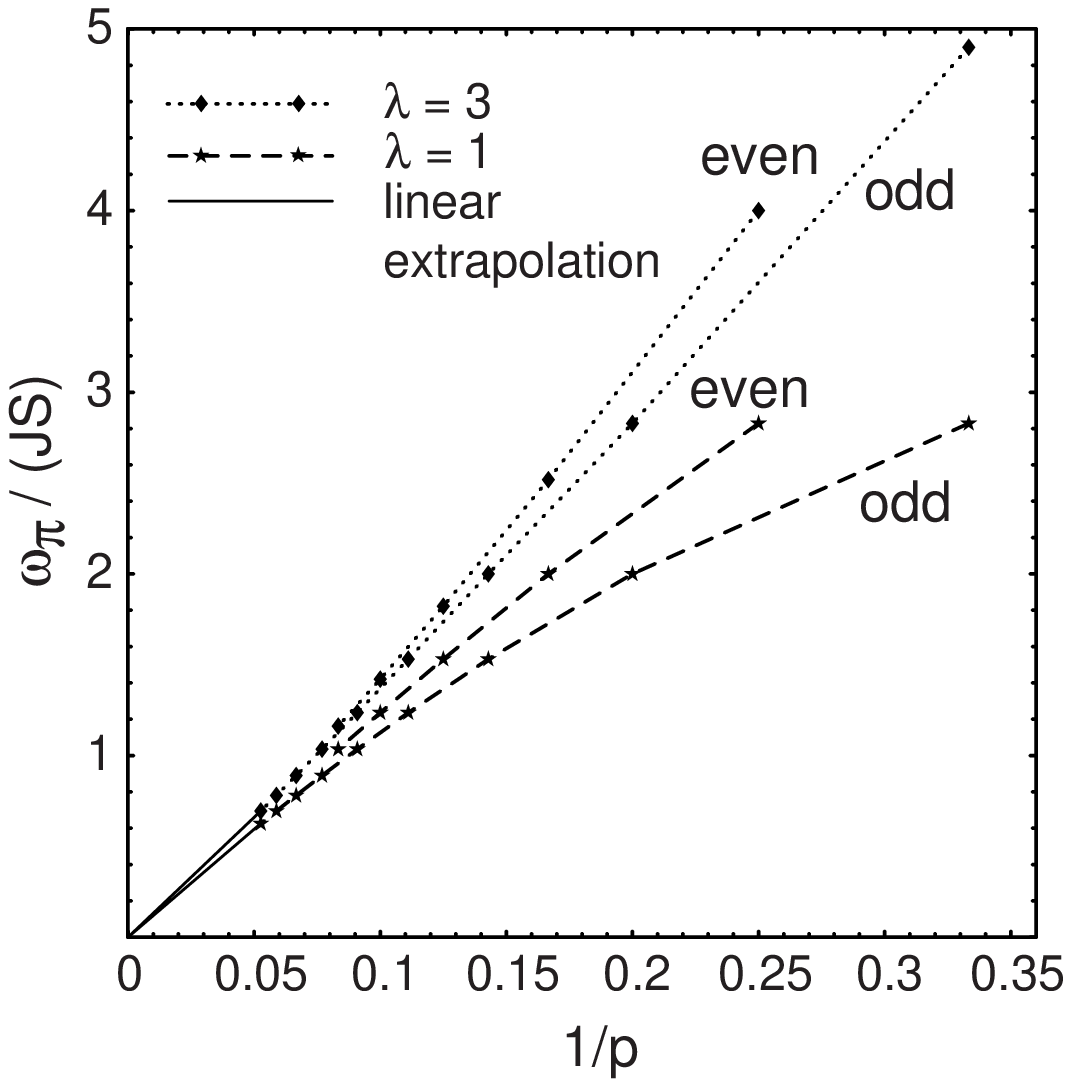,width=0.46\linewidth}
\caption{Resonance frequency $\w_\pi$  for diagonal stripes as a 
  function of $\lambda$ for different spacings $p$ (left) and as a
  function of $1/p$ for different couplings $\lambda$ (right; lines
  are a guide to the eye distinguishing even and odd $p$).}
\label{fig.d_w}
\end{figure}

With increasing coupling $\lambda$ the resonance energy increases.  In
contrast to vertical stripes, the resonance energy remains finite in
the limit $\lambda\to 0$ for even $p$ where it arises from an optical
band.  (cf.  Fig.~\ref{fig.d_w}).  Like for the vertical case,
$\w_\pi$ decreases with increasing stripe spacing, for $p\gg1$
according to $\w_\pi \propto 1/p$.  Since resonance comes from
different bands for even an odd $p$ the $\pi$, $\w_\pi$ is a
nonmonotonous function of $p$.  For this reason, $\w_\pi(p)$ is
plotted in Fig.~\ref{fig.d_w} separately for the two cases.

\section{Discussion}
\label{sec.disc}

We now discuss our findings in comparison to experimental data on the
spin dynamics, which are obtained predominantly from neutron
scattering.  As a result of this comparison we wish to advocate that
the simple stripe model provides a fair account of the spin dynamics
at not too low energies.  At very low energies, spin gaps may occur,
e.g., due to spin anisotropies (as in non-superconducting
LNO\cite{Nakajima+93} and LCO\cite{Keimer+93}), due to the coupling of
spins to the superconducting order parameter (as in superconducting
cuprates, see below), or simply due to the absence of
antiferromagnetic order (for too small $\lambda$).  Our model could
straightforwardly be generalized to account for the first origin.  The
inclusion of superconductivity would require a major extension.

In Table \ref{tab.1} we have collected basic parameters for various
undoped compounds setting the fundamental physical scales.  In Table
\ref{tab.2} spin dynamics data for specific stripe structures are
compiled.

\begin{table}[htbp]
  \begin{center}
    \begin{tabular}{l||c|c|c|c|c|c}
      material & 
      \# layers &
      S        & 
      $a$ &
      $J$ &
      $v_\AF$ &
      Refs. 
      \\\hline\hline
      LNO     & 
      1 &
      1 & 
      3.8 \AA &
      30 meV&
      0.32 eV \AA &
      \onlinecite{Sugai+90,Yamada+91,Nakajima+93} 
      \\\hline
      LCO     & 
      1 &
      $\frac 12$ & 
      3.8 \AA &
      135 meV &
      0.85 eV \AA &
      \onlinecite{Lyons+88,Hayden+90}
      \\\hline
      YBCO     & 
      2 &
      $\frac 12$ & 
      3.9 \AA &
      125 meV &
      n/a &
      \onlinecite{Sugai+90}
      \\\hline
      BSCCO    & 
      2 &
      $\frac 12$ & 
      3.8 \AA &
      140 meV &
      n/a &
      \onlinecite{Sugai+90}
    \end{tabular}
  \end{center}
  \caption{Basic parameters of the undoped parent compounds: 
    number of layers in the crystalline unit cell, 
    spin, nearest-neighbor spin spacing, 
    nearest-neighbor antiferromagnetic exchange coupling, 
    and spin-wave velocity.}
  \label{tab.1}
\end{table}

\begin{table}[htbp]
  \begin{center}
    \begin{tabular}{l||c|c|c|c|c}
      material & 
      $T_c$ & 
      $p$ & 
      $\omega_\pi$ & 
      $\omega_\gap$ &
      Refs.
      \\\hline\hline
      LSNO     & 
      0K &
      3 (d) & 
      80 meV &
      $\leq 28$ meV &
      \onlinecite{Bourges+02}
      \\\hline
      LSCO     & 
      $\approx 38$K &
      4 (v) & 
      n/a &
      3.5 meV &
      \onlinecite{Yamada+95,Petit+97,Lake+99}
      \\\hline
      LSCO     & 
      10K &
      6 (v) & 
      25 meV &
      $\leq 1.1$ meV &
      \onlinecite{Petit+97}
      \\\hline
      LSCO     & 
      0K &
      $\approx$ 43 (d) & 
      7 meV &
      0 meV &
      \onlinecite{Matsuda+00}
      \\\hline
      YBCO     & 
      90K & 
      5 (v) & 
      41 meV &
      28 meV &
      \onlinecite{Rossat-Mignod+91,Mook+93,Bourges+00} 
      \\\hline
      YBCO     & 
      63K & 
      n/a & 
      35 meV &
      28 meV &
      \onlinecite{Dai+96} 
      \\\hline
      YBCO     & 
      59K & 
      n/a & 
      26 meV &
      16 meV &
      \onlinecite{Rossat-Mignod+91} 
      \\\hline
      YBCO     & 
      39K & 
      8 (v)& 
      23 meV &
      10 meV &
      \onlinecite{Mook+02}
      \\\hline
      BSCCO    & 
      91K& 
      n/a & 
      43 meV&
      n/a&
      \onlinecite{Fong+99,He+01}
      \\\hline
      BSCCO    & 
      83K& 
      n/a & 
      38 meV&
      n/a&
      \onlinecite{He+01}
    \end{tabular}
  \end{center}
  \caption{Spin dynamics data for different materials at various doping levels 
    characterized by the critical temperature $\Tc$, 
    stripe period $p$ and orientation (diagonal/vertical), 
    resonance frequency $\w_\pi$, and gap frequency $\w_\gap$.}
  \label{tab.2}
\end{table}

\subsection{LSNO}

We start the comparison with LSNO which displays diagonal stripes and
where integer values of $p$ are particularly
stable\cite{Chen+93,Cheong+94} due to a lock-in of the stripes into
the atomic structure.  In this material, static stripes (i.e. stripes
that are visible down to $\w=0$) are seen at wave vectors $\bQ_\nu$.
\cite{Hayden+92} For $p=3$, the spin dynamics at higher energies has
been measured in detail.\cite{Bourges+02} Similar data are also
available for noninteger $p$, e.g.  $p=3.75$.\cite{Tranquada+97}

Experiments\cite{Yamada+91,Nakajima+93} on \textit{undoped} material
are in agreement with 2D spin-wave theory for the antiferromagnet with
$J \approx 30$ meV.  This exchange coupling corresponds to an
isotropic spin-wave velocity $v_\AF={\sqrt 8} S J a =0.32$
eV{\AA}\cite{Nakajima+93} since $S=1$ and $a \approx 3.8${\AA}.  This
agreement is reasonably good over a wide energy range $\w \gtrsim 30$
meV up to the band edge at $\w \approx 125$ meV, at low energies $\w
\lesssim 15$ meV deviations (gaps) appear\cite{Nakajima+93} due to a
uniaxial spin anisotropy and weak interlayer couplings.

The spin dynamics of the stripe system was examined for $p=3.75$ due
to oxygen doping\cite{Tranquada+97} as well as for $p=3$ with Sr
doping.\cite{Bourges+02} In the first case, a reduced velocity
$v_\parallel \approx 0.6 v_\AF$ was found in direction parallel to the
stripes, $v_\perp$ was not resolved.  In the second case, the velocity
was measured in both directions and found to be remarkably isotropic
and close to the value of the undoped system: $v_\parallel \approx
0.30$ eV{\AA} and $v_\perp \approx 0.35$ eV{\AA}. The overall shape of
the magnon dispersion was sinusoidal with an upper edge at $\w_\pi
\approx 80$ meV.

In our theory, this sinusoidal shape for $p=3$ is well reproduced
[compare Fig.~\ref{fig.d_HH} and Eq.~(\ref{sinu})].  The ratio
$\w_\pi/(JS) \approx 2.7$ is consistent with $\lambda \approx 0.9$.
For this value of $\lambda$, $v_\perp \approx v_\AF$ and $v_\parallel
\approx 0.67 v_\AF$.  Although we find $v_\parallel$ to be smaller
than in Ref.  \onlinecite{Bourges+02}, the overall agreement is very
satisfying and provides strong support for our case that the spin
dynamics can be well understood from a stripe model.  Small
quantitative deviations may be attributed to the simplicity of our
model using only two types of exchange couplings.

Remarkably, $\lambda \approx 0.9$ implies that the spin exchange
across a stripe is \textit{not much smaller} than within an AFM
domain.  It is important to keep in mind that $\lambda$ \textit{must
  not} be too small to preserve magnetic order.  A quantum Monte Carlo
analysis\cite{Matsumoto+01} of coupled two-leg ladders ($S=1$)
indicates a quantum phase transition into a disordered state at
$\lambda \approx 0.011$.  Below this value, stripe order would be
destroyed by quantum fluctuations.

Within our approach we can estimate also the two-magnon signal
accessible by Raman spectroscopy.  We may compare our single-magnon
density $\rho(\w)$ to the two-magnon scattering intensity at frequency
$2\w$. Certainly, this can be made only on a qualitative level, since
$\rho$ was calculated neglecting weight factors (which would change
the shape of spectra but not the frequency of resonances) and because
linear spin-wave theory does not include interactions between magnons.
Nevertheless, it is instructive to compare the outcome from our model
for the diagonal case $p=3$ with an experiment by Blumberg \textit{et
  al.}\cite{Blumberg+98} on LSNO.  In this experiment, two magnetic
resonances are observed at $\w \approx 4.6 J$ and $\w \approx 3 J$.
For $\lambda \approx 0.9$ we expect a singularity in the single-magnon
density at $\w \approx 2.7 JS$ (see Fig.~\ref{fig.zust}), which would
correspond to a two-magnon resonance at $\w \approx 5.4 JS$.  If
corrections due to magnon interactions are modest, the resonance of
the theory could be idebtified with the upper experimental one. Then
the resonance at the lower frequency cannot be understood.  On the
other hand, for $\lambda$ not too close to $1$ the single-band
structure for $p=3$ would lead to several well-separated extrema but
contradict the above determination of $\lambda$.  In particular, for
$\lambda <1$, the additional resonance lies above $\w_\pi$ since it
arises from extrema close to the upper band edge and there is only a
saddle-point at $\bk_\AF$.  This apparent contradiction might be
resolved if either interaction corrections are large, additional
exchange interaction are important, or the lower experimental
resonance is of different origin.

\subsection{Cuprates}

In the present study, we \textit{assume} the presence of charge
stripes and evaluate the spin dynamics for a simple model.  The
question of why stripes are formed and how stripe formation is related
to superconductivity therefore cannot be addressed.  In particular,
the simple spin-only model misses the coupling of spin fluctuations to
the superconducting order parameter.  Consequently, our analysis
misses the opening of a spin gap due to superconductivity.  Therefore,
the spin dynamics for $\omega<\omega_\gap$ is masked by
superconductivity (see Table \ref{tab.2}).  Nevertheless, one can
expect the stripe-like spin dynamics to remain visible in
superconducting samples for $\omega>\omega_\gap$.

Such a gap has been observed in experiments on LSCO (e.g.,
$\omega_\gap \simeq 3.5$ meV near optimal doping with
$\Tc=38.5$K;\cite{Yamada+95,Petit+97,Lake+99} a gap smaller than $1.1$
meV for underdoped samples with $\Tc=12$K and
$\Tc=25$K\cite{Petit+97}) and on YBCO (e.g., $\omega_\gap \simeq 10$
meV for a highly underdoped material with $\Tc=39$K;\cite{Mook+02}
$\omega_\gap \simeq 16$ meV for a moderately underdoped material with
$\Tc=59$K;\cite{Rossat-Mignod+91} $\omega_\gap \simeq 30$ meV for near
optimal doping with $\Tc=89$K\cite{Bourges+00}).  For YBCO there is
evidence\cite{Dai+01} for a proportionality between $\omega_\gap
\approx 3.8 \Tc$ which is not far away from the BCS weak-coupling
limit with $\omega_\gap = 3.52 \Tc$.

Furthermore -- and more importantly in the present context -- there is
evidence for such a (rough) proportionality not only between $\Tc$ and
$\omega_\gap$ but also between $\Tc$ and $\omega_\pi$ ($\omega_\pi
\simeq 5 \Tc$ for underdoped YBCO,\cite{Fong+97,Dai+99} $\omega_\pi
\simeq 5.4 \Tc$ for under- and overdoped BSCCO\cite{He+01}).  From our
theory, we expect $\w_\pi$ to be roughly inversely proportional to
$p$, see Eq.~(\ref{v}) and Fig.~\ref{fig.d_w}.  At low doping $p$
should be inversely proportional to the doping level ($x+2\delta$),
i.e., $\w_\pi$ should be proportional to the doping level.  Such a
relation was found in a previous theoretical study of the Hubbard
model,\cite{Demler+95} where it was attributed to a particle-particle
collective mode.  Although our approach is technically much less
involved, it provides an alternative explanation which is not in
contradiction with the previous one since stripe order itself can be
considered as a collective phenomenon that can be derived from the
Hubbard model.\cite{Zaanen+89,Schulz89}

On the other hand, at larger doping there is no simple relation
between the doping level and $p$.  In YBCO, for example, the
charge-transfer mechanism between the CuO$_2$ plane and the CuO chains
interferes.  In LSCO it is well documented that $p$ saturates at $p=4$
for $x \gtrsim 0.12$.\cite{Yamada+98} Beyond that point (which
corresponds to optimum doping\cite{Yamada+98}), additional holes may
populate the antiferromagnetic domains without affecting their period
$p$.  However, these excess holes may suppress the antiferromagnetic
exchange coupling in analogy to holes in the spin-glass phase (Ref.
\onlinecite{Tranquada+97} reports the corresponding suppression of the
spin-wave velocity).  Hence, the effective $J$ and, consequently, also
$\w_\pi$ may shrink with overdoping as seen in experiments on
BSCCO.\cite{He+01}

For LSCO, so far no direct evidence for a $\pi$ resonance has been
found.  This could be simply because the resonance intensity is
expected to be only $\sim 10$\% of the total magnetic
scattering.\cite{Bourges99} However, if the $\pi$ resonance -- in the
sense of a merger of the incommensurate signals -- can be attributed
to magnons in stripes which are particularly well established for
LSCO, one definitely should expect such a resonance.  For underdoped
LSCO ($p=6$, $\Tc=25$K) there is evidence for $\omega_\pi=25$ meV
(where incommensurate response becomes commensurate).\cite{Petit+97} A
similar signal was observed at even lower doping in the spin-glass
phase ($\omega_\pi=7$ meV for $p \approx 43$).\cite{Matsuda+00}

Like for LSNO, we may use the values of $J$, $p$, and $\w_\pi$ to
estimate $\lambda$ for the cuprates.  For YBCO with $J=125$ meV,
$p=5$, and $\w_\pi=41$ meV,\cite{Bourges+00,Dai+01} we obtain $\lambda
\sim 0.07$ from the left panel of Fig.~\ref{fig.v_w}.  If we take
$J=135$ meV and $\w_\pi=25$ meV for LSCO with $p=6$,\cite{Petit+97} an
even smaller value $\lambda \sim 0.04$ is found.

>From this result we may predict where the resonance $\w_\pi$ should be
expected in LSCO near optimal doping ($p=4$). For $J=135$ meV and
$\lambda=0.04-0.07$ we find $\w_\pi\approx 40-52$ meV. While the
resulting values for $\w_\pi$ have a certain spread, they suggest that
the resonance frequency should be at least as large as in optimally
doped YBCO.

In the experiments known to us, the considered energy range was simply
too small to detect the resonance for optimally doped LSCO: $\omega
\lesssim 6$ meV in \onlinecite{Yamada+95}, $\omega \lesssim 10$ meV in
Refs.~\onlinecite{Yamada+98,Lee+00}, $\omega \leq 16$ meV in Ref.
\onlinecite{Lake+99}.  However, from pulsed neutron scattering
evidence has been found for a broad peak in the momentum-integrated
susceptibility between 40 and 70 meV\cite{Yamada+95a,Arai+96} which
could be ascribed to the $\pi$ resonance.

Apparently, $\lambda$ seems to be significantly smaller in the
cuprates than in the nickelates.  At the same time, $S$ is smaller
(although $J$ is larger).  Therefore one may wonder whether static
magnetic stripe order is already destroyed by quantum fluctuations
without invoking competing orders leading to a gap.  For $S=\frac 12$
the coupling needs to satisfy $\lambda \gtrsim 0.3$ to stabilize spin
order for $p=3$,\cite{Tworzydlo+99,Matsumoto+01} while for $p=4$ a
finite $\lambda>0$ is sufficient.\cite{Tworzydlo+99} For $p=5$ (as for
every odd $p$) one again expects a finite critical $\lambda$.  If the
intersptripe coupling is below this value, the presence of a spin gap
can be understood also within the spin-only model.

\subsection{Conclusion}

In summary, we find that the spin fluctuations of stripes can provide
a simple and valuable description of the dynamics observed in
high-$\Tc$ compounds and related materials.  Already our minimalistic
spin-only model provides an accurate account of experiments on LSNO
and possibly also a unifying framework for incommensurate response and
the $\pi$ resonance in the cuprates.  While such a framework has been
suggested recently,\cite{Batista+01} it is analyzed and evaluated here
for the most transparent case of integer periods $p$.  Our results
unravel the evolution of the band structure with $p$ for diagonal and
vertical stripe configurations.  Likewise, we have explicitly
determined the dependence of characteristic spin-wave velocities and
of the resonance frequency on $p$ and $\lambda$.  Thereby, we
postulate that the $\pi$ resonance reflects the magnon frequency
$\w_\pi$ of the lowest lying band with nonvanishing weight.  In
particular, $\w_\pi$ was found to be roughly inversely proportional to
$p$ in agreement with experiments.

Hopefully, future experiments can provide more direct evidence for the
$\pi$ resonance also in LSCO.  This would also relax the controversial
question, whether spin excitations in LSCO and YBCO are
analogous\cite{Mook+98} or not.\cite{Bourges+00} If stripe magnons
indeed explain the spin dynamics at intermediate energies, as we
expect, they would provide a unifying framework for understanding the
spin dynamics above the gap scale.  Then the stripe physics would be
also of great importance as basement for superconductivity as
low-energy phenomenon.

Naturally, several aspects remain unexplained by our minimalistic
theory.  For example, our model cannot be expected to explain why the
magnetic incommensurability disappears at $\Tc$ in
YBCO\cite{Bourges+00} while charge order is visible up to
300K.\cite{Mook+02} Probably this is a question to the stripe-forming
mechanism and to a possible coupling between the order parameters for
stripe order and superconductivity.  In LSCO, the vicinity of soft
phonons and structural instabilities may help to stabilize stripes at
temperatures above the superconducting transition.

For future studies it would be interesting to include effects of the
bilayer coupling present in YBCO and BSCCO, of the weak 3D coupling
present in all materials, as well as spin anisotropy, more complicated
spin interactions (e.g.  four-spin cyclic
exchange\cite{Sugai+90,Muller-Hartmann+02}), excitations beyond spin
waves (e.g. double-spin excitations\cite{Blumberg+98}), mobility of
spins, and effects of disorder, to name just a few.

\begin{acknowledgments}
  We gratefully acknowledge helpful discussions with M. Braden, M.
  Gr\"uninger, B.  Keimer, E. M\"uller-Hartmann, T.  Nattermann, and
  G. S. Uhrig.  This project was supported financially by Deutsche
  Forschungsgemeinschaft (SFB608).
\end{acknowledgments}


\end{document}